\documentclass[3p]{elsarticle}
\usepackage[utf8]{inputenc}
\usepackage[english]{babel}    
\usepackage{amsmath}
\usepackage{amssymb}
\usepackage{graphicx}
\usepackage{bm}
\usepackage{booktabs}
\usepackage{tabularx}
\usepackage[table,x11names]{xcolor}
\usepackage[normalem]{ulem}

\newcommand{\op}[1]{\mathbf{#1}}
\newcommand{\im}{\mathrm i}
\definecolor{rottengreen}{rgb}{0, 0.49, 0.0}
\definecolor{dunkelgrau}{rgb}{0.4, 0.4, 0.4}

\usepackage{lineno,hyperref}
\hypersetup{colorlinks=true}
\usepackage{tcolorbox}
\modulolinenumbers[5]
\journal{Journal of \LaTeX\ Templates}
\begin{document}
\begin{frontmatter}
\title{Exceptional points in the thermoacoustic spectrum}
%
%
%
%
%
%
%
\author[Berlin]{Georg A.\ Mensah }
\author[Cambridge]{Luca Magri\corref{correspondingauthor}}
\ead{lm547@cam.ac.uk}
\author[Munich]{Camilo F.\ Silva}
\author[Trondheim]{Philip E.\ Buschmann}
\author[Trondheim,Berlin]{Jonas P.\ Moeck}
\cortext[correspondingauthor]{Corresponding author}
\address[Berlin]{Institut für Strömungsmechanik und Technische Akustik \\ Technische Universität Berlin \\ Berlin, Germany}
\address[Cambridge]{Engineering Department \\ University of Cambridge \\ Cambridge, UK}
\address[Munich]{Professur für Thermofluiddynamik \\ Technische Universität München \\ Munich, Germany}
\address[Trondheim]{Department of Energy and Process Engineering\\ Norwegian University of Science and Technology\\ Trondheim, Norway}
\begin{abstract}
Exceptional points are found in the spectrum of a prototypical thermoacoustic system as the parameters of the flame transfer function are varied. At these points, two eigenvalues and the associated eigenfunctions coalesce. The system's sensitivity to changes in the parameters becomes infinite. Two eigenvalue branches collide at the exceptional point as the interaction index is increased. One branch originates from a purely acoustic mode, whereas the other branch originates from an intrinsic thermoacoustic mode. The existence of exceptional points in thermoacoustic systems has implications for physical understanding, computing, modeling and control.
\end{abstract}
\end{frontmatter}
\section{Introduction}
At exceptional points (EPs), at least two eigenvalues and the associated eigenfunctions coalesce, and the eigenvalue sensitivity with respect to changes in the parameters becomes infinite~\cite{Kato1980,Heiss2012}. Interesting physical phenomena associated with EPs appear across various disciplines from quantum mechanics through optics and acoustics~\cite{Heiss2012,Ding2016,Achilleos2017}. 
To the best of the authors' knowledge, the role of exceptional points has not yet been explored in thermoacoustic systems, although points in the parameter space with infinite sensitivity were discussed in a recent review article~\cite{Juniper2018}. In this letter, we show that these points in the thermoacoustic spectrum are exceptional, and that they can be found in a generic thermoacoustic system when two real parameters are varied.
\subsection{Thermoacoustic instabilities}
Thermoacoustic instabilities are a major challenge for the reliable operation of many technical combustion systems, as reviewed by~\cite{Juniper2018} and references therein.  
For most practical applications with low-Mach number combustion, thermoacoustic phenomena can be modelled by an inhomogeneous Helmholtz equation, which reads
\begin{equation}\label{Helmholtz 00}
 \nabla\cdot\left( \bar{c}^{2} \nabla\hat{p} \right) + \omega^{2}\hat{p} = -\im\omega(\gamma - 1) \hat{\dot{q}}, 
\end{equation}
where $\omega$ is the complex frequency, $\bar{c}$ is the mean speed of sound, $\mathrm{i}^2=-1$, and $\gamma$ is the heat-capacity ratio. 
$\hat{p}$ and $\hat{\dot{q}}$ are the Fourier-transformed fluctuations\footnote{e.g., a fluctuation evolves as $\hat{(\cdot)}\exp\left(\mathrm{i}\omega t\right)$.} of acoustic pressure and heat release rate, respectively. Quantities are non-dimensionalized with a characteristic length, speed of sound, and density. The heat release rate fluctuation is commonly related to a velocity fluctuation at a reference position by a time-delay model~\cite{Juniper2018}, i.e.\ $-\im\omega(\gamma - 1) \hat{\dot{q}} = n\exp(-\im\omega\tau)\left.\nabla \hat p\right|_{x_\mathrm{ref}}$, where the parameters $n$ and $\tau$ are the interaction index and the time delay, respectively. 
The thermoacoustic stability problem is generally non-Hermitian because of the flame response term and dissipative boundary conditions. 
%
%
On numerical discretization or travelling-wave decomposition~\cite{Juniper2018}, 
thermoacoustic stability is governed by a nonlinear eigenvalue problem~\cite{Magri2016a,Guttel2017} 
\begin{align}
\label{eq:L_problemDef}
\op L(\omega;\bm \varepsilon) \bm{\hat p} = 0,  
\end{align}
where the vector $\bm \varepsilon\in\mathbb R^M$ contains $M$ parameters related to, for example, the mean speed of sound, the geometry, the flame response, and the boundary conditions. $\op{L}\in\mathbb C^{N\times N}$ is an analytic function of $\bm \varepsilon$ and $\omega$ in some subdomain of $\mathbb R^M\times\mathbb C$, where $N$ is the number of degrees of freedom of the discretized equations. 
For a given $\bm \varepsilon$, the stability of the linear system is characterized by the eigenvalues $\omega=\omega_r + \mathrm{i}\omega_i$, where $\omega_r\in\mathbb{R}$ is the angular frequency and $-\omega_i\in\mathbb{R}$ is the growth rate of the linear oscillation. 
With this convention, the system is linearly stable if $\omega_i>0$.  
The associated thermoacoustic mode shapes are provided by the eigenvectors $\bm{\hat{p}}\in\mathbb{C}^N$. 
\subsection{Eigenvalue classification}
Eigenvalues can be classified according to their algebraic and geometric multiplicities, $a$ and $g$. The algebraic multiplicity is the eigenvalue's multiplicity as a root of the dispersion relation, whereas the geometric multiplicity is the dimension of the associated eigenspace, i.e.\ the number of linearly independent eigenvectors. 
An eigenvalue of~\eqref{eq:L_problemDef} can be either semi-simple, when $a=g$; or defective, when $a>g$. For the special case $a=g=1$ an eigenvalue is called simple. Semi-simple eigenvalues with $g>1$ and defective eigenvalues are referred to as degenerate eigenvalues. Defective eigenvalues that are branch-point singularities in the parameter space are called exceptional points (EPs). 
On the one hand, eigenvalues of single-flame longitudinal thermoacoustic systems are typically simple~\cite{Juniper2018,Silva2018}. On the other hand, systems with discrete rotational symmetry, such as annular and can-annular combustors, feature semi-simple degenerate eigenvalues~\cite{Magri2016a,Mensah2016}, with fewer simple eigenvalues. 
\subsection{Sensitivity at an exceptional point}
Mathematically, in the neighborhood of an EP, the eigenvalue has a perturbation expansion in fractional powers of the parameter (Section II-2.2 in \cite{Kato1980}), also known as Puiseux series. 
At an EP with $a=2$ (hence $g=1$), which is assumed in the remainder of this letter, the change of the eigenvalue due to a perturbation to the $i$-th parameter, $\varepsilon_i$, reads 
\begin{align}
\omega = \omega_\mathrm{EP} + \omega_1 \sqrt{\varepsilon_i-\varepsilon_{i,\mathrm{EP}}} + O(\varepsilon_i-\varepsilon_{i,\mathrm{EP}})\qquad\varepsilon_i \to\varepsilon_{i,\mathrm{EP}},
\end{align}
where $\omega_1$ is a constant. 
Thus, the first-order sensitivity 
$\partial \omega/\partial  \varepsilon_i|_{\omega_\mathrm{EP},\bm\varepsilon_\mathrm{EP}}$ 
with respect to any parameter, $\varepsilon_i$, is infinite\footnote{This is in contrast with the semi-simple case, in which the first-order sensitivity is finite (Theorem II-2.3 in~\cite{Kato1980}).}~\cite{Heiss2012} because $(\omega-\omega_{\mathrm{EP}})/(\varepsilon_i-\varepsilon_{i,\mathrm{EP}})\rightarrow\infty$ as $\varepsilon_i\rightarrow\varepsilon_{i,\mathrm{EP}}$. An equivalent expansion holds for the eigenfunction at the EP. 
\subsection{Calculation of exceptional points in thermoacoustics}
We consider a thermoacoustic system with an $n$--$\tau$ flame model and calculate EPs as $n$ and $\tau$ are varied. 
The eigenvalues are the roots of the dispersion relation   
\begin{align} \label{eq:D=zero}
D(\omega; n ,\tau)=0,  
\end{align}
where $D(\omega; n ,\tau)\equiv\det \left[\op L(\omega; n,\tau)\right]$ is the characteristic function, which is transcendental and analytic in $\omega$ in some subdomain of the complex plane. 
For an eigenvalue to have $a=2$, \eqref{eq:D=zero} must be satisfied with the two following conditions 
\begin{align} 
\frac{\partial D}{\partial \omega}(\omega;n,\tau)=0, \label{eq:degenCon}\\
\frac{\partial^2 D}{\partial \omega^2}(\omega;n,\tau)\not=0.  \label{eq:degenCon1}
\end{align}  
The solution of the two complex-valued equations \eqref{eq:D=zero} and \eqref{eq:degenCon} is the set of parameters ($n_\mathrm{EP}$, $\tau_\mathrm{EP}$) and the defective eigenvalue $\omega_\mathrm{EP}$. Equations \eqref{eq:D=zero} and \eqref{eq:degenCon} would also be satisfied for degenerate semi-simple eigenvalues, such as those found in systems with rotational symmetry. However, in systems without symmetry, which we consider here, degenerate eigenvalues are generically defective \cite{Seyranian2005}. 
The defective eigenvalue has algebraic multiplicity two, but there is only one associated eigenvector $\bm{\hat p}_\mathrm{EP}$. 
\section{A prototypical time-delayed thermoacoustic system\label{sec:results}}
We consider a prototypical thermoacoustic system, which contains the essential physical mechanisms of many thermoacoustic systems~\cite{Juniper2018}. We assume that 
(i) the frequency of the oscillation is smaller than the cut-off frequency of the duct, i.e.\ only plane acoustic waves propagate; 
(ii) the duct has a sound hard end at the upstream boundary (zero acoustic pressure gradient) and an open end at the downstream boundary (acoustic pressure node); 
(iii) the flame is compact, i.e.\ it imposes a discontinuity in the mean temperature and acts as a point source for the acoustic field. 
The flame is located at the non-dimensional location $x_\mathrm{flm}=0.6$; the non-dimensional reference position, at which the acoustic velocity drives the flame, is $x_\mathrm{ref}=0.5$; the ratio of the speeds of sound between the hot and cold side is 2. The reference quantities for non-dimensionalization are the length  of the duct and the speed of sound / density of the cold side.
The characteristic function for this classical thermoacoustic problem reads 
\begin{align}
&D(\omega;n,\tau)=\nonumber\\ &n \exp(-\im\omega\tau) \sin(\omega x_\mathrm{ref} ) \sin\left((x_\mathrm{flm}-1) \frac{\omega}{2}\right) +
 \sin(x_\mathrm{flm} \omega ) \sin\left((x_\mathrm{flm}-1)\frac{\omega}{2}\right) +2\cos(x_\mathrm{flm} \omega ) \cos\left((x_\mathrm{flm}-1) \frac{\omega}{2}\right).
\end{align}
Table \ref{tab:points} lists the acoustic mode ($n=0$) and the EPs found in the vicinity of it by solving Eqs.~\eqref{eq:D=zero} and \eqref{eq:degenCon}. The EPs approach the acoustic eigenvalue as $\tau$ increases, while the magnitude of the associated interaction index $n$ decreases. Section~\ref{sec:PS} discusses the eigenvalue and eigenvector sensitivity in the vicinity of the EP \#1.a. The results for the other EPs in Tab.~\ref{tab:points} are qualitatively similar (result not shown).
\begin{table}
\centering
\caption{Acoustic eigenvalue (in grey) and some close-by exceptional points. The parameters $n$ and $\tau$ are given to ten decimal places. With this precision, two eigenvalues are found to be identical up to four decimal places.
\label{tab:points}}
\begin{tabularx}{0.57\columnwidth}{l c c c c }
\toprule
\# &$n$ & $\tau$ &$\operatorname{real}(\omega)$ & $\operatorname{imag}(\omega)$  \\
\midrule
\rowcolor{gray}\color{white}1&\color{white} $0.0$ &\color{white}  - &\color{white}$2.2273$ &\color{white} $0.0000$ \\
1.a& $ 1.0332748434 $ & $ 0.9545158731 $ &$ 2.5967 $ & $ 0.7160 $  \\
1.b& $ 0.3397183716 $ & $ 3.6122221467 $&$ 2.2766 $ & $ 0.2615 $ \\
1.c& $ 0.1367125262 $ & $ 9.2027775468 $&$ 2.2357 $ & $ 0.1076 $ \\
1.d & $ -0.0939726697 $ & $ 13.4231684179 $&$ 2.2313 $ & $ 0.0741 $
\\\bottomrule
\end{tabularx}
\end{table}
\section{Exceptional points in the thermoacoustic spectrum \label{sec:PS}} 
Because the algebraic multiplicity of the EPs considered here is $a=2$, two eigenvalues will be found in the vicinity of the defective eigenvalue as the parameters depart from the exceptional point. 
In combination with the extreme sensitivity close to the EP, the numerical computation of EPs is therefore challenging for algorithms based on fixed-point iteration, such as those commonly used in thermoacoustic analyses. In the present work, a global contour-integral-based method proposed by Beyn~\cite{Guttel2017} is used. This method provides all the eigenvalues within a given circle in the complex plane, even if they are defective. The integration circle has been centered at the defective eigenvalue \#1.a with unit radius. This circle encloses the acoustic eigenvalue $\omega_\mathrm{ac}\approx 2.2273+0\im$ (mode \#1).
\begin{figure}
\centering
 \includegraphics[width=\textwidth]{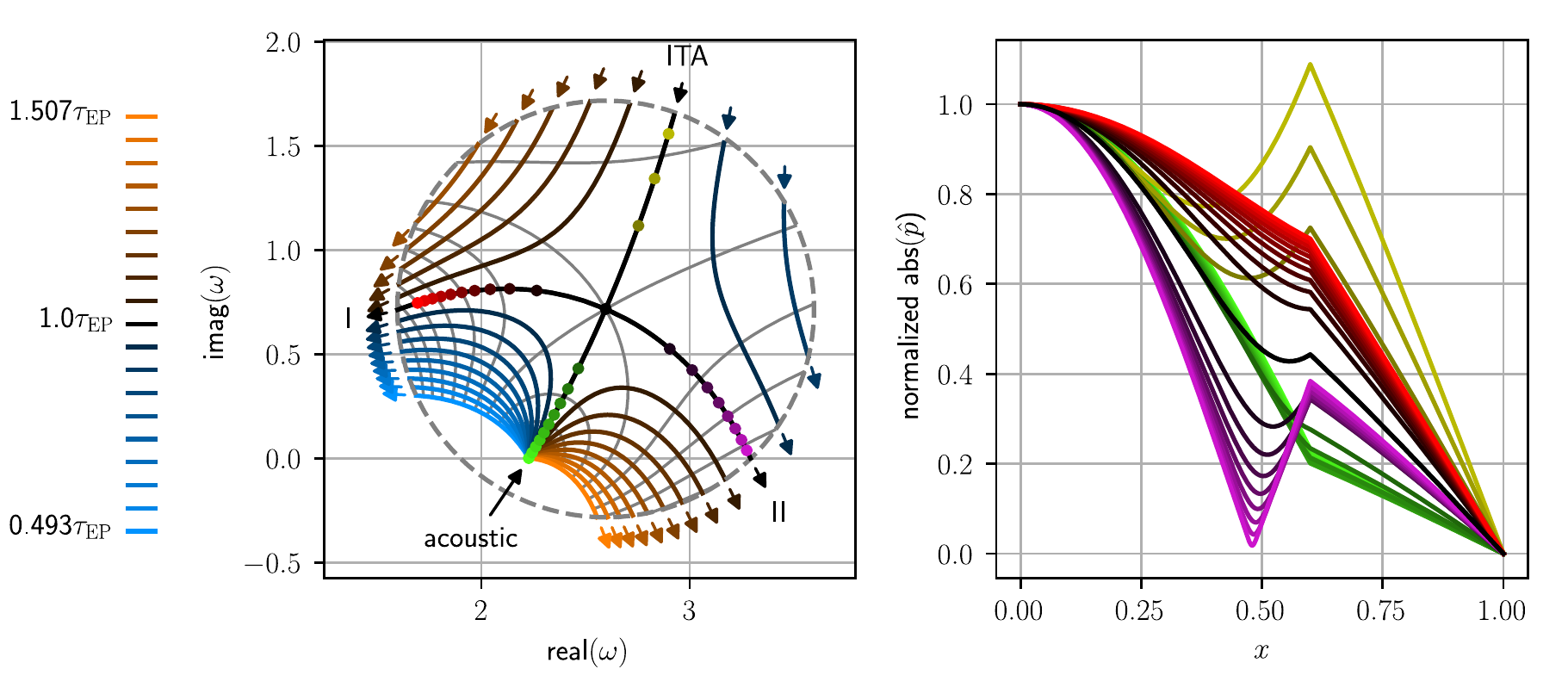}
 \caption{
 Left: eigenvalue trajectories when $n$ is varied from 0 to 3 times its exceptional value (mode \#1.a in Table~\ref{tab:points}). 
 Blue lines are for $\tau<\tau_\mathrm{EP}$ while orange lines are for $\tau>\tau_\mathrm{EP}$. $\tau$ varies equidistantly between $\tau_\mathrm{EP}\pm 0.2 \frac{2\pi}{\operatorname{real}(\omega_\mathrm{EP})}$. The darker the shading, the closer the values are to the exceptional point. The black lines indicate the trajectories for $\tau=\tau_\mathrm{EP}$; their intersection indicates the EP. The colored arrows indicate the direction of increasing $n$. The thin grey lines highlight solutions for constant $n$. The markers on the black line depict values of $n$ ranging equidistantly from $0$ to $2n_\mathrm{EP}$. Only solutions inside the circle are shown. Right: the eigenvectors corresponding to the markers on the exceptional trajectories. Acoustic branch in green, intrinsic thermoacoustic (ITA) branch in yellow, exceptional branch I in red, exceptional branch II in purple, EP in black.
\label{fig:paths}}
\end{figure}
Figure~\ref{fig:paths} (left panel) shows the eigenvalue trajectories in the vicinity of EP \#1.a, which are parametrized by the interaction index $n$ for different levels of $\tau$. 
When $\tau=\tau_\mathrm{EP}$, while the interaction index $n$ is varied from zero to $3n_{\mathrm{EP}}$, two eigenvalue trajectories (black lines) approach each other, coalesce at $n=n_{\mathrm{EP}}$ and diverge eventually. 
At the EP, the eigenvalue trajectories cross each other, i.e., they coalesce. 
This is a manifestation of the branch-point singularity, which implies infinite parameter sensitivity.
The acoustic eigenvalue $\omega_\mathrm{ac}$ is the starting point of the trajectory labeled `acoustic'. It is neutrally stable because the system without flame is conservative. 
 The trajectory coming from the opposite direction starts far away from the circle with a large positive imaginary part, which, in contrast to the acoustic mode, corresponds to a highly damped mode. 
As observed in~\cite{Silva2018} (and references therein), the physical origin of this trajectory is an intrinsic thermoacoustic (ITA) mode, which, for $n\ll1$, is highly damped and independent of the geometry.
Almost all the trajectories in the vicinity of the EP, thus, originate from either an acoustic mode or an intrinsic mode. 
The exceptions to this rule are the branches I and II, which are of mixed type, thus, they cannot be unambiguously traced back further than the EP. The investigation of these branches is left for future work. 
 The large parameter sensitivity becomes apparent when considering eigenvalue trajectories that do not pass across the EP. The curvature and spread of these lines show that the parameter sensitivity becomes larger as $n$ and $\tau$ approach the EPs. 
Figure~\ref{fig:paths} (right panel) shows the absolute value of the eigenvectors for different points along the exceptional branch ($\tau=\tau_\mathrm{EP}$). 
Because the exceptional point is a defective eigenvalue, the two mode shapes collapse at $n= n_\mathrm{EP}$, i.e. $g=1$ (black line). 
A small perturbation to $n$ significantly changes the mode shape around the EP. 
\section{Discussion}
Exceptional points in the spectrum of a prototypical thermoacoustic system are found and investigated for the first time, to the best of the authors' knowledge. 
In contrast to semi-simple degenerate eigenvalues, which are found in the thermoacoustic analysis of annular combustors and have finite sensitivity, EPs do not stem from a geometric symmetry of the system.
These points are branch-point singularities in the parameter space. They have fundamental and practical implications for thermoacoustic stability. 
\begin{itemize}
\item Physics:  Exceptional points occur when two eigenvalue trajectories with different physical nature collide. One trajectory originates from an acoustic mode,  and the other trajectory originates from an intrinsic thermoacoustic mode.
\item Numerical methods: Iterative methods based on fixed point algorithms, which are commonly used in thermoacoustic stability analysis, do not seem to be robust in the vicinity of exceptional points. 
A contour-integration-based approach~\cite{Guttel2017} facilitates robust computations of the thermoacoustic spectrum.
\item Modeling and control: The large sensitivity at an EP may help design new control schemes to mitigate thermoacoustic instabilities with small changes in the design variables. The appropriate expansion at the EP, which can be used to calculate sensitivities to the system's parameters for passive control, is in fractional powers of the parameters. Robust control schemes will be necessary around exceptional points because small uncertainties in the parameters are exceedingly amplified. 
\end{itemize}
Future research will be aimed at establishing the universality of EPs in thermoacoustic systems, investigate the role of EPs in systems with discrete rotational symmetry, and exploit the properties of EPs, e.g.\ the large sensitivity to parameters, for control of instabilities.
\section*{Acknowledgements}
L.M. gratefully acknowledges support from the Royal Academy of Engineering Research fellowship. 
\section*{References}

\begin{thebibliography}{10}
\expandafter\ifx\csname url\endcsname\relax
  \def\url#1{\texttt{#1}}\fi
\expandafter\ifx\csname urlprefix\endcsname\relax\def\urlprefix{URL }\fi
\expandafter\ifx\csname href\endcsname\relax
  \def\href#1#2{#2} \def\path#1{#1}\fi

\bibitem{Kato1980}
T.~Kato, {Perturbation theory for linear operators}, 2nd Edition, Springer
  Berlin / Heidelberg, New York, 1980.

\bibitem{Heiss2012}
W.~D. Heiss, \href{http://stacks.iop.org/1751-8121/45/i=44/a=444016}{The
  physics of exceptional points}, Journal of Physics A: Mathematical and
  Theoretical 45~(44) (2012) 444016.
\newline\urlprefix\url{http://stacks.iop.org/1751-8121/45/i=44/a=444016}

\bibitem{Ding2016}
K.~Ding, G.~Ma, M.~Xiao, Z.~Q. Zhang, C.~T. Chan,
  \href{https://link.aps.org/doi/10.1103/PhysRevX.6.021007}{Emergence,
  coalescence, and topological properties of multiple exceptional points and
  their experimental realization}, Physical Review X 6 (2016) 021007.
\newblock \href {http://dx.doi.org/10.1103/PhysRevX.6.021007}
  {\path{doi:10.1103/PhysRevX.6.021007}}.
\newline\urlprefix\url{https://link.aps.org/doi/10.1103/PhysRevX.6.021007}

\bibitem{Achilleos2017}
V.~Achilleos, G.~Teocharis, O.~Richoux, V.~Pagneux, Non-{H}ermitian acoustic
  metamaterials: {R}ole of exceptional points in sound absorption, Physical
  Review B 95 (2017) 144303.
\newblock \href {http://dx.doi.org/10.1103/PhysRevB.95.144303}
  {\path{doi:10.1103/PhysRevB.95.144303}}.

\bibitem{Juniper2018}
M.~P. Juniper, R.~Sujith,
  \href{https://doi.org/10.1146/annurev-fluid-122316-045125}{Sensitivity and
  nonlinearity of thermoacoustic oscillations}, Annual Review of Fluid
  Mechanics 50~(1) (2018) 661--689.
\newblock \href {http://dx.doi.org/10.1146/annurev-fluid-122316-045125}
  {\path{doi:10.1146/annurev-fluid-122316-045125}}.
\newline\urlprefix\url{https://doi.org/10.1146/annurev-fluid-122316-045125}

\bibitem{Magri2016a}
L.~Magri, M.~Bauerheim, M.~P. Juniper,
  \href{http://www.sciencedirect.com/science/article/pii/S0021999116303278}{Stability
  analysis of thermo-acoustic nonlinear eigenproblems in annular combustors.
  {Part I. Sensitivity}}, Journal of Computational Physics 325 (2016) 395 --
  410.
\newblock \href {http://dx.doi.org/https://doi.org/10.1016/j.jcp.2016.07.032}
  {\path{doi:https://doi.org/10.1016/j.jcp.2016.07.032}}.
\newline\urlprefix\url{http://www.sciencedirect.com/science/article/pii/S0021999116303278}

\bibitem{Guttel2017}
S.~Güttel, F.~Tisseur, The nonlinear eigenvalue problem, Acta Numerica 26
  (2017) 1–94.
\newblock \href {http://dx.doi.org/10.1017/S0962492917000034}
  {\path{doi:10.1017/S0962492917000034}}.

\bibitem{Silva2018}
C.~F. Silva, K.~J. Yong, L.~Magri, Thermoacoustic modes of quasi-1{D}
  combustors in the region of marginal stability, Proceedings of the ASME 2018
  Turbomachinery Technical Conference \& Exposition, GT2018-76921 (2018) 1--12.

\bibitem{Mensah2016}
G.~Mensah, G.~Campa, J.~Moeck, Efficient computation of thermoacoustic modes in
  industrial annular combustion chambers based on {B}loch-wave theory, Journal
  of Engineering for Gas Turbines and Power 138 (2016) 081502 (7 pages).

\bibitem{Seyranian2005}
A.~P. Seyranian, O.~N. Kirillov, A.~A. Mailybaev, Coupling of eigenvalues of
  complex matrices at diabolic and exceptional points, J. Phys. A: Math. Gen.
  38 (2005) 1723--1740.

\end{thebibliography}
\end{document}